# Classification of Human Monkeypox Disease Using Deep Learning Models and Attention Mechanisms

Md. Enamul Haque
*Department of Computer Science and Engineering*
*United International University*
mhaque182055@bscse.uiu.ac.bd

Md. Rayhan Ahmed
*Department of Computer Science and Engineering*
*United International University*
rayhan@cse.uiu.ac.bd

Razia Sultana Nila
*Department of Computer Science and Engineering*
*United International University*
rnila182041@bscse.uiu.ac.bd

Salekul Islam
*Department of Computer Science and Engineering*
*United International University*
salekul@cse.uiu.ac.bd

*Abstract*—As the world is still trying to rebuild from the destruction caused by the widespread reach of the COVID-19 virus, and the recent alarming surge of human monkeypox outbreaks in numerous countries threatens to become a new global pandemic too. Human monkeypox disease symptoms are quite similar to chickenpox, and measles classic symptoms, with very intricate differences such as skin blisters, which come in diverse forms. Various deep-learning methods have shown promising performance in the image-based diagnosis of Covid-19, tumor cell, and skin disease classification tasks. In this paper, we try to integrate deep transfer learning-based methods, along with a convolutional block attention module (CBAM) to focus on the relevant portion of feature maps to conduct an image-based classification of human monkeypox disease. We implement five deep learning models—VGG19, Xception, DenseNet121, EfficientNetB3, and MobileNetV2 along with integrated channel and spatial attention mechanisms and perform a comparative analysis among them. An architecture consisting of Xception-CBAM-Dense layers performed better than the other models at classifying monkeypox and other diseases with a validation accuracy of 83.89%.

## I. INTRODUCTION

According to World Health Organization (WHO) [1], Monkeypox is a rare disease caused by the monkeypox virus. It can spread from animals to people and is transferable between people and the environment since it is a viral zoonotic infection. The monkeypox outbreak became the most talked-about topic in the world after the COVID-19 pandemic was over. Up to this point, a total of 72,874 confirmed cases have been discovered, including 28 deaths across 12 locations[2]. Fever, headache, muscle aches, back discomfort, low energy, and swollen lymph nodes are the most typical signs of monkeypox, which can also be followed by the appearance of a rash[1]. The face, palms of the hands, soles of the feet, groin, genital, and/or anal regions might all be affected by the rash[1]. The slight distinction between monkeypox and Others (measles, chickenpox) is inflammation and rash on the body, which is difficult to detect by human eyes except by the Polymerase Chain Reaction (PCR) Test.

It is exceedingly difficult for medical professionals to make an accurate diagnosis of this disease quickly. Additionally, it is difficult to find the PCR test used to detect the monkeypox virus. However, prompt diagnosis of this illness is necessary to stop the spread by adhering to other WHO-provided guidelines and the first step of preventing monkeypox, which calls for keeping isolation and a distance from other people, and so on.

In recent research, it is clear that deep learning is far more sophisticated than classical machine learning models[3] like KNN, Random Forest classifier, and so on. However, they remain machines that solve the categorization problem by requiring additional domain knowledge and human involvement. Additionally, traditional machine learning models are typically only effective for the tasks for which they were created. Therefore, the traditional machine learning approach may not be as effective in this subject.

Due to the superior learning capabilities of deep learning models notably the variants of convolutional neural networks (CNNs), have recently transformed various fields of medical science [4][5]. These deep networks can analyze images in several layers, automatically identify significant features, and learn to discover the best representations for particular tasks when trained with a huge amount of data. Also transferring learning models helps to tackle the problem of the enormous dataset demand.

Transfer learning is another method that is frequently employed when there is a lack of data. With this technique, a CNN model that has already been trained on a big dataset (such as ImageNet) is used to transfer its knowledge to a separate, much smaller dataset for context-specific learning. As a result, employing the transfer learning method will speed up, simplify, and improve our performance compared to other traditional ways.

In this study, we used the Monkeypox Skin Lesion Dataset (MSLD) [6] and a few transfer-learning-based models with the CBAM attention mechanism. Used transfer learning model for comparing is - VGG19 [7], DenseNet121[8], Xception [9], EfficientNetB3 [10] and MobileNetV2 [11] and we were able to predict the enhanced train data with an accuracy of 69.86% for VGG19 and 78.27% for DenseNet121, 54.21% for EfficientNetB3 model, 74.07% for MobileNetV2 model and 79.90% for proposed Xception model.

The remaining papers are divided into the following sections, with section II covering the specifics of the related work with our paper, section III states the methodology and related materials that are required for the study, section IV states experimental details like dataset description, hyperparameter tuning information and evaluation metrics details and with result analysis.

## II. RELATED WORK

Recently many researchers have explored diverse deep learning algorithms for the detection of human monkeypox disease. Ali et al. [6] Incorporated VGG-16, ResNet50, and InceptionV3 pre-trained models for the classification of monkeypox and other diseases (chickenpox, measles) on the Monkeypox Skin Lesion Dataset (MSLD) which was developed by the authors. The dataset contains images of skin lesions in two categories. One is Monkeypox and Others (Chickenpox, Measles). The overall accuracy ResNet50 achieved is 82.96(±4.57%), and VGG-16 achieved 81.48(±6.87%). The author ensembled these three models and resulting in an accuracy of 79.26 (±1.05%).

In another study, Islam et al. [12] examined the performance of the monkeypox classification over pre-trained ResNet50, DenseNet121, Inception-V3, SqueezeNet, MnasNet-A1, MobileNet-V2, and ShuffleNet-V2 models. Based on ShuffleNet-V2 model obtained a maximum of 79% accuracy in comparison to the other models used, with 85% precision, demonstrating the possibility of employing AI models. The Monkeypox Skin Image Dataset 2022 was also a dataset the authors developed and used

Ahsan et al. [13] created the Monekypox2022 image dataset first before proposing a modified VGG16 model using two separate studies. The suggested modified VGG16 has an accuracy of 97±1.8% (AUC: 97.2) for study one and 88±0.8% (AUC: 0.867) for study two in identifying monkeypox patients. Additionally, they talked about their model's feature extraction and prediction using Local Interpretable Model-Agnostic Explanations (LIME).

Sitaula et al. [14] analyzed and evaluated thirteen different pre-trained deep-learning models for the detection of the monkeypox virus. DenseNet-169 and Xception performed better than others on the Monkeypox2022 image dataset. Based on performance DenseNet-169 and Xception were ensembled. Additionally, the ensembled method offers accuracy (87.13%), precision (85.44%), recall (85.47%), and F1-score (85.40%).

Sahin et al. [15] used some pre-trained deep learning models on the MSLD dataset and discovered that MobileNetV2 and EfficientNetB0 model performed well. They then constructed a mobile application that can categorize monkeypox disease after converting the entire model into a TensorFlow lite model embedding with metadata.

## III. METHODOLOGY

To test the feasibility of classifying human monkeypox in this analysis, we picked five pre-trained deep learning models through the transfer learning process, ranging from heavy-weight architectures like VGG-19 and Xception to lightweight models like DenseNet121, MobileNetV2, and EffcientNetB3. Figure 1 depicts the overall pipeline of the training procedure for those models. Due to the lack of images in the utilized dataset, we selected deep models over a variety of sizes to examine the impact of training sample size. All pre-trained models are customized using the same method. Moreover, to focus the network on more relevant feature maps with we incorporate the convolutional block attention module.

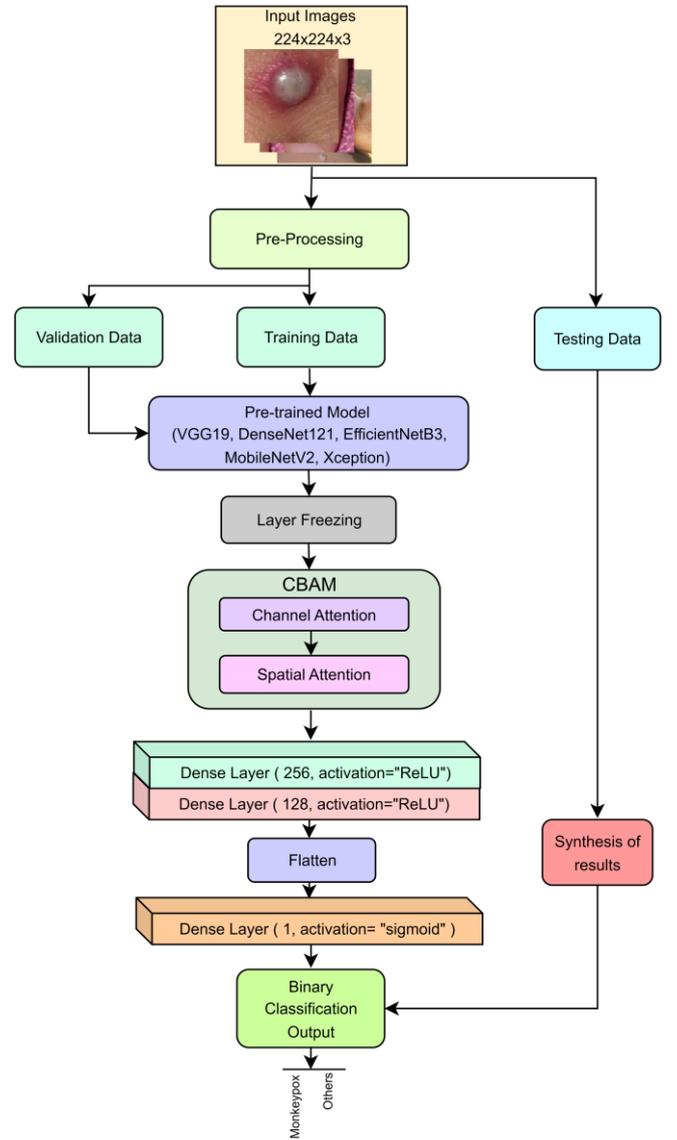

*Figure 1. Proposed Architecture*

At first Human Monkeypox Skin Lesion Dataset have been prepared for training by resizing the image to a resolution of 224 x 224 x 3. To be noted, we have frozen every layer except the last two layers of all the pre-trained architectures. To utilize the most relevant parts of the generated feature maps we utilized the convolutional block attention module (CBAM), proposed by Woo et al. [16]. CBAM contains two sequential attention-based mechanisms - Channel attention followed by Spatial attention. The taken output from the CBAM was passed through two dense layers of 256 and 128 neurons with ReLU activation function. For the conversion of the layer into 1D to feed the last layer, we used Flatten and fed the output into the last dense layer with sigmoid activation to perform binary classification. Finally, validation data were used to synthesize the outcome from the training data.

CBAM is an adaptive image refinement module, which is sequentially applied in the channel and spatial dimensions and is used to concentrate more on major features of the images. The overall summarized attention process is:

$$F' = M_c(F) \otimes F \quad \ldots\ldots\ldots\ldots (1)$$

$$F'' = M_s(F') \otimes F' \quad \ldots\ldots\ldots (2)$$

Where $F$ = feature map ($F \in R^{C \times H \times W}$ = input), 1D channel attention map $M_c \in R^{C \times 1 \times 1}$ and 2D spatial attention map $M_s \in R^{1 \times H \times W}$ and $F''$ is the final refined output, and $\otimes$ denotes element-wise multiplication.

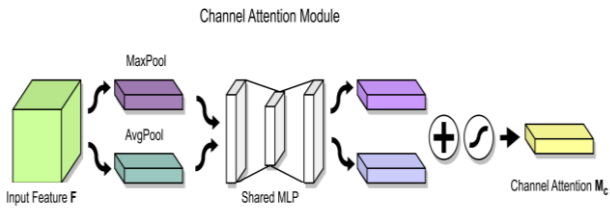

*Figure 2. Design of the Channel attention module* [16].

Channel attention [16] as illustrated in Figure 2. focuses on the inter-channel relationship of features and the reduction of channel redundancy. The features of average-pooling and max-pooling were combined to aggregate the two separate spatial descriptors $F^c_{avg}$ and $F^c_{max}$.

The channel attention map $M_c \in R^{C \times 1 \times 1}$ is obtained by feeding these descriptors into a shared network made up of a multi-layer perceptron (MLP) with a single hidden layer, after which the output feature vectors are subjected to a merging process. Overall, the calculation is:

$$M_c(F) = \sigma(MLP(AvgPool(F)) + MLP(MaxPool(F)))$$

$$= \sigma(W_1(W_0(F^c_{avg})) + W_1(W_0(F^c_{max}))) \quad \ldots\ldots\ldots (3)$$

Here, $\sigma$ = sigmoid function, $W_0, W_1$ = MLP weights.

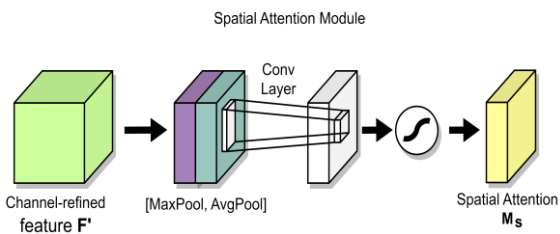

*Figure 3. Design of the spatial attention module* [16].

Spatial attention [16] as demonstrated in Figure 3. focus on the informative part or content information of an input feature in the spatial position. On the input feature, the process applies average pooling and maximum pooling, which are then concatenated to produce the feature descriptor. After that, a convolutional layer was used to create the $M_s(F) \in R^{H \times W}$ spatial attention map. A conventional convolutional layer concatenates the results of the pooling operations, $F^s_{avg} \in R^{1 \times H \times W}$ and $F^s_{max} \in R^{1 \times H \times W}$ to create a 2D spatial attention map. The calculations are defined as follows:

$$M_s(F) = \sigma(f^{7 \times 7}([AvgPool(F); MaxPool(F)]))$$

$$= \sigma(f^{7 \times 7}([F^s_{avg}; F^s_{max}])) \quad \ldots\ldots\ldots (4)$$

Here, $\sigma$ = Sigmoid function, $f^{7 \times 7}$ = Convolutional operation, and filter size of 7x7.

## IV. EXPERIMENTS AND ANALYSIS

### A) Dataset

To classify Monkeypox among chickenpox, measles, and monkeypox we used Monkeypox Skin Lesion Dataset (MSLD) [6]. This image dataset includes a total of 228 original images of measles and chickenpox. Some of the images are demonstrated in fig. 3 and fig. 4. The dataset includes both original images and augmented folded images (Train, Test, Validation) with a proportion of 70:10:20. We used the data from the augmented Train folder for training, which contains 2142 images in 2 classes (Monkeypox, Others). Here, the training and validation images from the dataset were augmented while the test set contained only the original images.

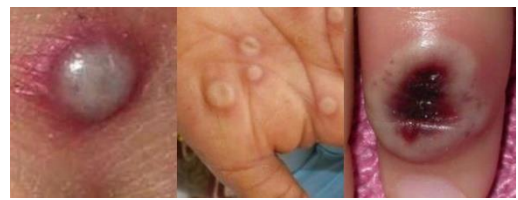

*Figure 4. Monkeypox skin lesion sample images.*

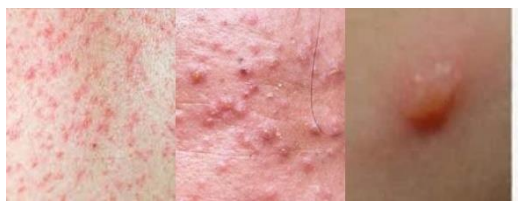

*Figure 5. Others (chickenpox, measles) skin lesion sample images.*

### B) Optimization and hyper-parameter tuning

In this paper, we have used various hyper-parameter to ensure a good performance from the model. In order to reduce the losses, we used adam optimizer with a learning rate of 0.001. We also employed binary_crossentropy that computes the cross-entropy loss between true labels and predicted labels with accuracy metrics too. To get the best model from the execution we have employed Model Checkpoint with other required tunings for monitoring val_loss metrics to monitor the behavior of the accuracy of the model. For all experiments, the batch size was fixed at 32.

C) Evaluation Metrics

The most prominent statistical methods, including accuracy, precision, recall, and F1-score, are employed to measure and present the overall experimental outcome. They are defined as follows:

$$\text{Accuracy} = \frac{T_P + T_N}{T_P + T_N + F_P + F_N} \quad \quad (5)$$

$$\text{Precision} = \frac{T_P}{T_P + F_P} \quad \quad (6)$$

$$\text{Recall} = \frac{T_P}{T_N + F_P} \quad \quad (7)$$

$$\text{F1-Score} = 2 * \frac{\text{Precision} * \text{Recall}}{\text{Precision} + \text{Recall}} \quad \quad (8)$$

D) Result Analysis

For the classification of monkeypox among other similar diseases (Chickenpox, Measles) we have employed five pre-trained architecture models with attention mechanisms and with dense layers. It should be noted that the results presented in Table I are the average of performance across four different folds (4-fold cross-validation). The confusion matrix generated each of the four-fold is presented in fig. 6. We have stated the average accuracy, precision, recall, and f1-score of the VGG19, DenseNet121, EfficientNetB3, MobileNetV2, and Xception architectures along with CBAM and Dense layers for classification in Table I.

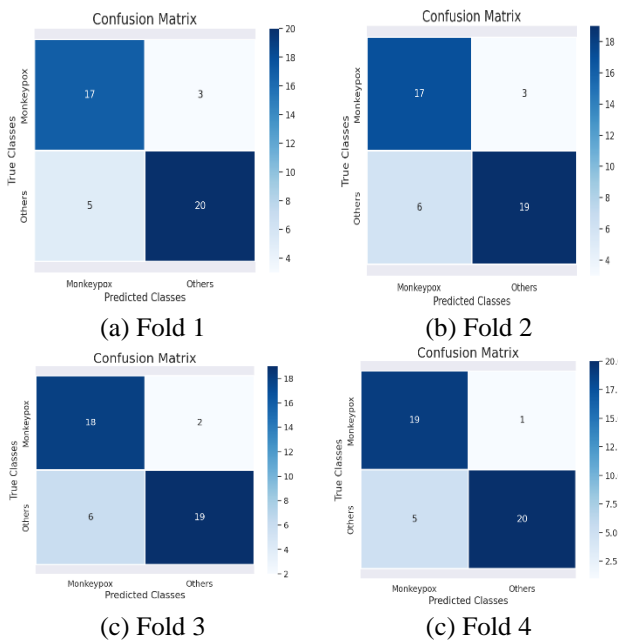

Figure 6. Generated confusion matrix from the 4-fold cross validation.

From Table I. we can observe that the Xception-CBAM-Dense architecture performed better than other investigated combinations with a validation accuracy, precision, recall and f1-score of 83.89%, 90.70%, 89.10%, and 90.11%, respectively. All the scores presented in Table I are validation scores. As for the value of evaluated metrics on the test dataset, the VGG19-CBAM-Dense architecture, Xception-CBAM-Dense architecture, DenseNet121-CBAM-Dense architecture, MobileNetV2-CBAM-Dense architecture, and lastly the EfficientNetB3-CBAM-Dense architecture achieved a test accuracy of 70.71%, 85.83%, 79.28%, 75.28%, and 84.37%, respectively.

TABLE I: A comparative summary of the acquired evaluation metric scores (validation) using different variants of the pre-trained models along with CBAM and Dense layers, and with other studies that used the same MSLD dataset (- denotes not mentioned).

| Model Architecture | Accuracy (%) | Precision (%) | Recall (%) | F1_Score (%) |
|---|---|---|---|---|
| VGG19-CBAM-Dense | 71.86 | 73.91 | 72.90 | 73.31 |
| **Xception-CBAM-Dense** | **83.89** | **90.70** | **89.10** | **90.11** |
| DenseNet121-CBAM-Dense | 78.27 | 80.49 | 76.31 | 81.56 |
| MobileNetV2-CBAM-Dense | 74.07 | 77.16 | 72.22 | 75.19 |
| EfficientNetB3-CBAM-Dense | 81.43 | 85.61 | 81.05 | 86.38 |
| Ensemble [6] | 79.26 | 84.00 | 79.00 | 81.00 |
| ShuffleNet [15] | 80.00 | - | - | - |

V. CONCLUSION

This research work presents a comparative analysis and preliminary feasibility study of five pre-trained deep learning algorithms along with a convolutional block attention module (CBAM) and dense layers for the classification of monkeypox disease from skin lesions by using the Monkeypox Skin Lesion Dataset (MSLD). Implementation of the CBAM module with channel and spatial attention module permits the proposed network to emphasize the major and effective feature maps from the image features and focus on inter-channel dependencies of the diseased regions of the image. The proposed model obtains reasonable classification performance with a validation accuracy of 83.89% using the Xception-CBAM-Dense layer architecture. The proposed study produced superior findings when compared to other studies using the same dataset. One of the major drawbacks to the proper training of the proposed model is the extremely small number of original images of monkeypox and other diseases in the dataset. In future work, we hope to design a more optimized model and train it on a larger dataset of human monkeypox disease images.